\documentclass[11pt,a4paper]{article}
\usepackage{anziamjedraft}
\usepackage{amssymb,amsfonts}

\title{Adventures in Invariant Theory}

\author{P.~D. Jarvis and J.~G. Sumner}
\address{School of Mathematics and Physics, University of Tasmania, 
Private Bag 37 GPO, Hobart Tas~7001, \textsc{Australia}.}
\mailto{Peter.Jarvis@utas.edu.au,Jeremy.Sumner@utas.edu.au}

\date{July 2013}

\begin{document}

\maketitle

\begin{abstract}
\noindent
We provide an introduction to enumerating and constructing
invariants of group representations via character methods. 
The problem is  contextualised via two case studies, arising from our recent work: 
entanglement invariants, for characterising the structure of state spaces for composite quantum systems; 
and Markov invariants, a robust alternative to parameter-estimation intensive methods of
statistical inference in molecular phylogenetics. 
\end{abstract}


\section{Introduction}
What can the pursuits of (\emph{i}) investigating quantum entanglement, via multicomponent wavefunctions, on the one hand, and (\emph{ii}) studying frequency array data in order to infer species evolution in molecular phylogenetics, on the other -- both hot topics in their respective fields -- possibly have to do with one another? Quite a lot, as it turns out -- as becomes clear, once the elegant connections with group representations and tensor analysis are made transparent. The following is an overview of some of the salient background, and a biassed selection of applications of invariant theory to the respective topics, arising from our recent work in both areas.
The results which we report here provide
novel instances of how group representation theory, and specifically classical invariant theory, can provide well-founded and useful tools for practitioners, in the realms of both quantum information, and mathematical biology. 

Given a group $G$ and a $G$-module $V$ (a space carrying a linear $G$ action, or representation), there is a standard construct ${\mathbb C}{[}V{]}$, the space of `polynomials in the components of the vectors in $V$'. Natural objects of special interest in this space are the `invariants', that is, functions $f(x)$ which are unchanged (up to scalar multiplication)\footnote{Of course, $\lambda_g$ must be a one-dimensional representation, $\lambda_g \lambda_h = \lambda_{gh}$, which for the cases studied here will be realized by various matrix determinants.} under the action of $G$, $f(g\!\cdot\! x) = \lambda_g f(x)$, and we would like to characterize the sub-ring of invariants, $I(V) := {\mathbb C}{[}V{]}^G$. In view of the grading of ${\mathbb C}{[}V{]}$ by degree, the coarsest characterization is the associated Molien series, $h(z) = \sum_0^\infty h_n z^n$ with $h_n = Dim({\mathbb C}{[}V{]}{}^G_n)$. In well-behaved cases, $I(V)$ has a regular structure (and is finitely generated), and $h(z)$ is a very pleasant rational polynomial. For $G$ semi-simple and compact, Molien's theorem \cite{molien1897invarianten} gives an integral representation of $h(z)$ via the Haar measure on $G$.
Knowledge of $h(z)$ and of a set of generators of $I(V)$ is generally important for applications. For example if $V$ is the adjoint representation, with $G$ semi-simple, Harish-Chandra's isomorphism states that $I(V)$ is a polynomial ring, whose generators are nothing but the fundamental (Casimir) invariants for the Lie algebra ${\mathfrak g}=L(G)$ of $G$. For a comprehensive introduction to the theory of representations and invariants of the classical groups see for example Goodman \cite{Goodman1998}. We now turn to our discussion of applications.
\section{Application I -- Quantum entanglement}
In nonrelativistic quantum mechanics with continuous variable systems, we work with the Schr\"{o}dinger representation, whose uniqueness is guaranteed by the celebrated Stone-von Neumann theorem. The $V$'s are thus various complex $L^2$ spaces and, for multi-partite systems, tensor products thereof.  However, for purely `spin' systems, where the state space is spanned by a finite set of eigenstates of some selected observable quantity, the Hilbert spaces are simply finite-dimensional complex vector spaces  $V \cong {\mathbb C}^N$. Our interest here is in composite systems with $K$ parts. In the context of quantum information, a subsystem with dimension $D$ is referred to as a `qu$D$it'. For $K$ qu$D$its, then, $N=K^D$. The simplest case occurs for $D=2$ (corresponding to spin-$\frac 12$, for example `up' or `down' electronic spin states in an atom) and we have $K$ `qubits', with $V$ the $K$-fold tensor product
${\mathbb C}^2 \otimes {\mathbb C}^2 \otimes  \cdots {\mathbb C}^2$ of dimension $N=2^K$.

The quantum state of the system as a whole is described as usual by a vector in the total space $V$, but we imagine experimenters Alice, Bob, Carol, $\cdots$, and Karl who are each able to access only one subsystem. In the oft-described scenario of `spooky-action-at-a-distance', Alice, Bob, Carol, $\cdots$, and Karl, despite remaining in their spatially separated labs, each manipulate their own subsystem independently, but observable outcomes between their measurements, and those in their colleagues' labs, are nonetheless not independent -- the properties of each subsystem's quantum state in this case are correlated with those of the other $K\!-\!1$ subsystems, and the overall state is described as `entangled'.

	%

One strategy available to each of Alice, Bob, Carol, $\cdots$, and Karl is simply to let his or her individual subsystem change under some time evolution, which can be engineered independently of the others. However, such local transformations do not affect the entanglement of the joint, $K$-party quantum state of the system as a whole: hence, any proposed numerical measure of entanglement must be invariant under appropriate symmetry transformations. Since standard time evolution of quantum states is represented by unitary operators, entanglement measures should therefore be invariant under the Cartesian product of $K$ unitary groups, each  acting on one experimenter's qu$D$it Hilbert space. In the qubit case, then, the symmetry group is just\footnote{More general procedures open to Alice, Bob, Carol, $\cdots$, and Karl involve various types of general quantum operations (measurements). For example, under reversible operations which succeed only with some probability less than one, the transformation group on each subsystem would be extended from $U(2)$ to $GL(2,{\mathbb C})$, and the group as a whole would become $\times^K GL(2,{\mathbb C})$. Of course, such local transformations do modify state entanglement, although numerical measures which are {bona fide} \emph{entanglement monotones} are defined to be \emph{nonincreasing} under such changes \cite{Vidal2000em}.}
$G = U(2)\times U(2) \times \cdots \times U(2)$ acting on the said $K$-fold tensor product space 
$V \cong \otimes^K {\mathbb C}^2$.

The invariants from $I(V)$ are perfectly suited to quantifying these local quantum effects, and are hence referred to as `local entanglement invariants'.
There is great interest in using these invariants to build complete entanglement measures \cite{Vidal2000em}, and the first problem is to characterize and evaluate the invariants in different situations. A famous case in point for tripartite entanglement ($K=3$) is the use of the Cayley hyperdeterminant, which is called the \emph{tangle} in the physics literature \cite{CoffmanKunduWootters2000de}. See \cite{HorodeckiEtAl:2009qe} for a recent review of the topic of quantum entanglement.

A less well studied case is that of so-called mixed states, where the system itself is described in a statistical sense (an ensemble of electrons, each of whose members is an electron described by a state vector which is an equal superposition of spin `up' and spin `down', is physically very different from an ensemble wherein, in 50\% of instances, the electron spin is `up', and in the other 50\%, the electron spin is `down'). 
The state is now specified by a density operator (a self-adjoint positive semidefinite linear operator on $V$ of unit trace), and hence transforms in the adjoint representation $\cong  V \otimes V^*$. Even just for $K=2$, that is for \emph{two} qubit mixed states, the structure of the invariant ring is quite rich, for example being considerably more complicated than the \emph{four} qubit pure state case \cite{VerStraeteEtAl2002fq9e}. The Molien series  \cite{GrasslEtAl1998cli,makhlin2002nlp,KingWelshJarvis2007}
\[
h(z) = \frac{1+z^4 + z^5 + 3z^6 + 2z^7+ 2z^8 +3z^9 +z^{10} + z^{11} + z^{15} }
{(1-z)(1-z^2)^3(1-z^3)^2 (1-z^4)^3 (1-z^6)}
\]
enumerates a plethora of primary and secondary invariants, whose precise role in the formulation of suitable entanglement measures is still not completely tied down \cite{GrasslEtAl1998cli,makhlin2002nlp}. 
%
\section{Application II -- Phylogenetics}
What of molecular phylogenetics? The simplest, so-called `general Markov model' of molecular evolution \cite{barry1987,semple2003phylogenetics} is given as follows. For a given set of $K$ species (`taxonomic units'), a probabilistic description of some set of $D$ observed characters is adopted. Models are constructed that describe the frequency of patterns derived from morphological features, or in molecular phylogenetics, from alignments of homologous nucleic acid sequences, nucleotide bases $\{A,C,G,T\}$, $D=4$; or of homologous proteins, amino acid residues $\{A,R,N,D,C,E,Q,G,H,I,L,K,M,F,P,S,T,W,Y,V\}$, $D=20$; or a variety of other molecular motifs or repeated units. These models are constructed by assuming molecular sequences evolving from a common ancestor via a Markov process, punctuated by speciation events. The data, corresponding to the observed frequencies, are taken as a sample of the probabilities on the basis that each site in the alignment independently follows an identical random process. These assumptions are contestible, but are well motivated by considerations of finding a balance between biological realism and statistical tractability. Contained within this model is the description of the evolution of the $K$ \emph{extant} species and their characters. 
This is a process whereby the $K\!$-way probability array, sampled by the pattern frequencies, evolves according to the tensor product of $K$ independent $D \times D$ Markov transition matrices. This scenario is analogous to the set-up of quantum entanglement described above, and algebraically it becomes an instance of the classical invariant theory problem, by extending the \emph{set} of Markov matrices to the smallest containing matrix \emph{group}. In the case of continuous-time models, this is no difficulty, as the matrices describing substitution rates between molecular units formally belong to the relevant matrix Lie algebra \cite{sumner2012}, and the Markov transition matrices are their matrix exponentials -- and are hence invertible. From this algebraic perspective, it also makes sense to work over $\mathbb{C}$ from the outset, and later examine stochastic parameter regions as required for applications. This will be elaborated upon for a specific example below.

The said $K$-fold tensor product module ${\mathbb C}^D \otimes {\mathbb C}^D \otimes  \cdots \otimes {\mathbb C}^D$ thus now transforms under $G=GL_1(D) \times GL_1(D) \times \cdots \times GL_1(D)$, where the non-reductive 
group\footnote{This group is thus the workhorse of Markov models, playing a role analogous to $GL(D)$, which Weyl in his book famously referred to as `her all-embracing majesty' amongst the classical groups.} $GL_1(D)$ is the Markov \emph{stochastic group} of invertible $D\!\times\!D$ unit row-sum matrices  \cite{johnson1985,mourad2004} ($GL_1(D)$ is of course a matrix subgroup of $GL(D)$, and is isomorphic to the affine group $A\!f\!\!f_{D\!-\!1}$ in one dimension lower; the \emph{doubly stochastic} group is the subgroup having unit row- \emph{and} column-sums, and is isomorphic to $GL(D\!-\!1)$ (see \S A below)). In this non-reductive case there is no Molien theorem, and no guarantee of the invariant ring even being finitely generated. However, there is no difficulty in counting one dimensional representations degree by degree in tensor powers, and indeed we have shown that a slightly modified version of the standard combinatorial results applies (see Appendix). In practical terms, this allows us to identify useful invariants for the purposes of phylogenetic inference. In this context we call such objects \emph{Markov invariants}.

One such quantity, the so-called `\texttt{logDet}', has in fact been known and used by phylogenetic practitioners for over two decades \cite{barry1987,lake1994,lockhart1994}. For the case of two taxa, the determinant function of the 2-fold phylogenetic tensor array (a polynomial of degree $D$) is certainly a one dimensional representation under the action of $GL(D) \times GL(D)$ itself, in fact transforming as $Det \otimes Det$, and thus necessarily an invariant of the Markov subgroup. Taking the (negative) log, and with the usual matrix relation $-\ln Det = -Tr \ln$, we recover the (negative) sum of the traces of the rate generators, multiplied by the evolved time. Modulo some care with the distribution of characters belonging to the presumed common ancestor of the two taxa, this can be taken as a measure of the total `evolutionary distance' between them, essentially the sum of all the individual rates changing characters into one another, multiplied by the time. The \texttt{logDet} can be recorded for all pairs of taxa, using marginalisations of the $K$-fold probability array, and thus leads to a robust `distance-based' method for phylogenetic inference. In fact,
Buneman's theorem \cite{buneman1971} guarantees reconstruction of a tree from a pairwise `metric' satisfying certain additional conditions.

Using our technical results, Markov invariants beyond the two-fold case are able to be counted and constructed, and it is an important matter of principle to investigate them. 
In data sets where the number of species $K$ is large, where the pairwise nature of \texttt{logDet} can lead to significant loss of evolutionary information, they
may also provide alternative or supplementary information to help with inference. In view of the pevious discussion of quantum entanglement, it turns out that for the case of binary characters ($D=2$), and three-fold arrays ($K=3$) or tripartite marginalisations of higher arity arrays, the Cayley hyperdeterminant (degree $n=4$) is precisely such a candidate \cite{sumner2005}, and we have identified analogous low-degree `tangles' for $D = 3$ and 4 \cite{sumner2006}.
For four taxa, $K=4$, and four characters, $D=4$, we have found a remarkable, symmetrical set of three degree-five, $n=5$, Markov invariants dubbed the `squangles' (\underline{s}tochastic \underline{qu}artet t\underline{angles})\cite{sumner2008,sumner2009,holland2013low}.  
A simple least squares analysis of their values \cite{holland2013low} allows a direct ranking of one of the three possible unrooted tree topologies for quartets\footnote{It is here that careful account of the stochastic parameter regime should be taken, as a crucial aspect of the least squares analysis requires certain inequalitites to hold.}. 
The squangles provide a low-parameter and statistically powerful way of resolving quartets based on the general Markov model \cite{holland2013low}, without any special assumptions about the types of rate matrices in the model, and independently of any recourse to pairwise distance measures. They are useful because many reconstruction methods for large trees build a `consensus tree' from some kind of ranking of quartet subtrees, where robust decisions at the quartet level are absolutely crucial. Further details are given in the appendix, \S A.

It must be noted that Markov invariants are in general distinct from the so-called `phylogenetic
invariants' \cite{CavenderFelsenstein1987}. These are polynomials that evaluate to zero for a subset of phylogenetic trees regardless of particular model 
parameters, and hence can serve in principle to discriminate trees and models. Their formal presentation
can be given in terms of algebraic geometry \cite{Lake1987,allman2004}. However, in contrast to Markov invariants which are 1-dimensional $G$-modules, phylogenetic invariants in general belong to high-dimensional $G$-modules \cite{allman2013,sumner:jarvis:2013cpi}.

Our Markov invariants are necessarily quite large objects -- they are polynomials of reasonably high degree in a significant number of variables. For example, the squangles are degree 5 polynomials in the components of a $4^4 = 256$-element array, and given their combinatorial origins, it is perhaps not surprising to find that they each have 66,744
terms\footnote{This is still $\ll O(256^5)$.}. However, once defined, there is no numerical problem with evaluations\footnote{Explicit forms for the squangles, together with \texttt{R} code for their evaluation, are available from the authors.} -- their utility is in their ability to syphon useful information out of the complexity of the data. As such they provide a viable alternative to parameter-estimation intensive phylogenetic methods, where massive likelihood optimisations are required, in order to make decisions about much more tightly specified models.

\paragraph{Acknowledgements}
The authors thank E Allman, D Ellinas, B Fauser, J Fern\'{a}ndez-S\'{a}nchez, B Holland, R King, J Rhodes, M Steel and A Taylor for helpful discussions and correspondence on this research. JGS acknowledges the support of the Australian Research Council grant DP0877447 for part of this work. PDJ acknowledges the support of the Australian-American Fulbright Commission for the award of a senior scholarship for part of this work.

\begin{appendix}
\section{Counting invariants: some character theorems}
The mathematical setting for both the study of entanglement measures for composite quantum systems, and of analogous quantities for the setting of phylogenetics, is that there is a model space
$V$ which is a $K$-fold tensor product, $V \cong {\mathbb C}^D\otimes {\mathbb C}^D\otimes \cdots 
\otimes  {\mathbb C}^D$. In the case of quantum mechanics the components of $V$ in some standard basis describe the state; for example in Dirac notation a pure state is a ket $|\Psi \rangle \in V$ of the form
$
|\Psi \rangle = \sum_{0}^{D\!-\!1} \Psi_{i_1 i_2 \cdots i_K} |i_1,i_2, \cdots, i_K \rangle
$
in the case of qu$D$its (see below for mixed states). In the phylogenetic case we simply have a $K$-way frequency array
${\{} P_{i_1 i_2 \cdots i_K} {\}}$ sampling the probability of a specific pattern, say ${i_1 i_2 \cdots i_K}$, where each $i_k \in {\{} A,C,G,T {\}}$ for nucleotide data, at a particular site 
in a simultaneous alignment of a given homologous sequence across all $K$ of the species under consideration. 

We focus attention on the linear action of the appropriate matrix group $G = G_1 \times G_2 \times \cdots \times G_K$ on $V$.
In the quantum qu$D$it case each local group $G_k$ is a copy of $U(D)$, but given the irreducibility of the fundamental representation, for polynomial representations the analysis can be done using the character theory of the complex group\footnote{This technical point is different from the previous observation about extending the analysis to allow local quantum operations and communication of these between parties.} $GL(D, {\mathbb C})$. This group is too large for the phylogenetic case, where the pattern frequency array $P$ evolves as $P \rightarrow P' := g \cdot P$, namely
\[
P' = M_1 \otimes M_2 \otimes \cdots \otimes M_K \cdot P
\]
where each $M_k$ belongs to the stochastic Markov group $GL_1(D,{\mathbb C})$ (the group of nonsingular complex unit row-sum $D \!\times\!D$ matrices). 

We compute the Molien series $h(z) = \sum_0^\infty h_n z^n$ for ${\mathbb C}{[} V{]}^G$ degree-by-degree using combinatorial methods based on classical character theory for $GL(D)$, adapted slightly for the stochastic case $GL_1(D)$, which we now describe. All evaluations are carried out using the group representation package ${}^\copyright \!$\texttt{Schur} \cite{SCHUR}. 

In terms of class parameters
(eigenvalues) $x_1,x_2,\cdots, x_D$ for a nonsingular matrix $M \in GL(D)$, the defining representation, the character is simply $Tr(M) =  x_1+ x_2+ \cdots + x_D$; the contragredient has character
$Tr(M^T{}^{-1}) =  x_1{}^{-1}+ x_2{}^{-1}+ \cdots + x_D{}^{-1}$. Irreducible polynomial and rational characters of $GL(D)$ are given in terms of the celebrated Schur functions \cite{Weyl1939,littlewood1940} denoted $s_\lambda(x)$, where $\lambda = (\lambda_1,\lambda_2,\cdots,\lambda_D)$, $\lambda_1 \ge \lambda_2 \ge \cdots \ge \lambda_D$, is an integer partition of at most $D$ nonzero parts. 
$\ell(\lambda)$, the length of the partition, is the index of the last nonzero entry (thus $\ell(\lambda)=D$ if $\lambda_D >0$). $|\lambda|$, the weight of the partition, is the sum $|\lambda|=\lambda_1+\lambda_2 + \cdots + \lambda_D$, and we write $\lambda \vdash |\lambda|$. For brevity we write the Schur function simply as $\{\lambda \}$ where the class parameters are understood. Thus the space $V$ as a representation of $G$ as a $K$-fold Cartesian product is endowed with the corresponding product of $K$ characters of the above defining representation of each local group, $\chi= \{1\} \cdot \{1\}\cdot \, \cdots \,  \cdot \{1\}$ in the quantum mechanical pure state and stochastic cases, and 
$\chi = (\{1\} \{\overline{1}\}) \! \cdot \! (\{1\} \{\overline{1}\})\!\cdot \, \cdots \, \cdot\!(\{1\} \{\overline{1}\})$ in the quantum mechanical mixed state case, where $\{1\}$ is the character of the defining representation, and $\{\overline{1}\}$ that of its contragredient.
The space of polynomials of degree $n$ in $\Psi$ or $P$, ${\mathbb C}{[}V{]}_n$, is a natural object of interest and by a standard result is isomorphic to the $n$-fold symmetrised tensor product $V \vee V \vee \cdots \vee V$, a specific case of a Schur functor: ${\mathbb S}_{\{n\}}(V)$. Its character is determined by the corresponding Schur function \emph{plethysm}, 
$\chi \underline{\otimes} \{n\}$, and the task at hand is to enumerate the one-dimensional representations occurring therein.

Before giving the relevant results it is necessary to note two further rules for combining Schur functions. The \emph{outer} Schur function product, is simply the pointwise product of Schur functions, arising from the character of a tensor product of two representations. Of importance here is the \emph{inner} Schur function product $\ast$ defined via the Frobenius mapping between Schur functions and irreducible characters of the symmetric group. We provide here only the definitions sufficient to state the required counting theorems in technical detail. For a more comprehensive, Hopf-algebraic setting for symmetric functions and characters of classical (and some non-classical) groups see \cite{FauserJarvis2003hl,FauserJarvisKing2006nbr}. 

Concretely, we introduce structure constants for inner products in the Schur function basis as follows:
\[
\{\lambda \} \ast \{ \mu \} = \sum_\nu g^\nu_{\lambda,\mu} \{\nu \}.
\]
For partitions $\lambda$, $\mu$ of equal weight\footnote{If 
$|\lambda| \ne |\mu|$ then $\{\lambda \} \ast \{ \mu \}=0$.}, $|\lambda| = |\mu|= n$, say, this expresses the reduction of a tensor product
of two representations of the symmetric group ${\mathfrak S}_n$ labelled by partitions $\lambda$, $\mu$. By associativity, we can extend the definition of the structure constants to $K$-fold inner products,
\[
\{\tau_1 \} \ast \{\tau_2 \} \ast \cdots \ast \{\tau_K \} = \sum_\nu g^\nu_{\tau_1, \tau_2, \cdots, \tau_K}
\{\nu \}.
\] 

\noindent
\textbf{Theorem: Counting invariants}
\begin{description}
\item[(a) Quantum pure states]\mbox{}\\
Let $D$ divide $n$, $n = rD$, and let $\tau$ be the partition $(r^D)$ (that is, with Ferrers diagram a rectangular array of $r$ columns of length $D$). Then 
\[
h_n = g^{(n)}_{\tau,\tau,\cdots,\tau}\quad \mbox{($K$-fold inner product)}.
\] 
If $D$ does not divide $n$, then $h_n =0$.
\item[(b) Quantum mixed states]\mbox{}\\ We have
\[
h_n = \sum_{|\tau|= n,\ell(\tau) \le D^2}
\left( \sum_{|\sigma|= n, \ell(\sigma) \le D} g^{\tau}_{\sigma,\sigma}\right)^{\!\!\!\!2}.
\]
\item[(c) Phylogenetic $K$-way pattern frequencies, general Markov model]
We have
\[
h_n = g^{(n)}_{\tau_1,\tau_2,\cdots,\tau_K}\quad \mbox{($K$-fold inner product)},
\] 
for each $\tau_k$ of the form $(r_k+s_k, r_k^{(D\!-\!1)})$ such that $n = r_kD+s_k$, $s_k \ge 0$.
\item[(d) Phylogenetic $K$-way pattern frequencies, doubly stochastic model]
We have 
\[
h_n = g^{(n)}_{\tau_1,\tau_2,\cdots,\tau_K}\quad \mbox{($K$-fold inner product)},
\] 
for each $\tau_k$ of the form $(r_k+s_k, r_k^{(D\!-\!2)},t_k)$ such that 
$n = r_k(D\!-\!1)+s_k+t_k$, $0\le t_k \le r_k$, $s_k \ge 0$.
\end{description}
\mbox{}\hfill $\Box$\\

As pointed out in the main text, the enumeration and identification of entanglement invariants, in the case of quantum systems, and Markov invariants, in the phylogenetic context, is of practical importance in characterising general properties of the systems under study -- in the quantum case, because they are by definition impervious to local unitary operations, and form the raw material for constructing interesting entanglement measures; and in the phylogenetic case, because they tend to be independent of how the specific Markov change model is parametrized, but nonetheless they can give information about the underlying tree. 

An example of identifying invariants is the case of the three squangle quantities. We find ${g^{(5)}}_{\tau\tau\tau\tau}=4$, where $\tau$ is the partition $(2,1^3)$ which is of course of dimension $4$ and irreducible in $GL(4)$, but indecomposable in $GL_1(4)$, as it contains a one-dimensional representation. One of the four linearly independent degree five candidates is discounted, because of algebraic dependence on lower degree invariants. Recourse to the appropiate quartet tree isotropy group \cite{sumner2009} reveals that one of the remaining three is not tree informative. Further, the situation with respect to the final two objects is expressed symmetrically in terms of the \emph{three} squangle quantities $Q_1$, $Q_2$, $Q_3$, which satisfy $Q_1+Q_2+Q_3=0$, as follows. For tree 1, for example $12|34$, we have on evaluation with stochastic parameters, $Q_1= 0$, and $-Q_3=Q_2 > 0$. This pattern recurs cyclically for the other two unrooted quartet trees: for tree 2, $13|24$, $Q_2=0$, whereas $-Q_1=Q_3 > 0$, and for tree 3, $14|23$, $Q_3=0$, and $-Q_2=Q_1>0$. As noted above, the (strict) inequalities entailed in the above evaluations are crucial for the validity of the least squares method for ranking quartet trees using squangles.

There are many more gems to be examined in hunting down Markov invariants for different models and subgroups \cite{jarvis:sumner:2012miesm,jarvis:sumner:2013missm}, with potential practical and theoretical interest. As one 
instance
of as-yet unexplored terrain, for $K=3$ we have evidence \cite{sumner2006a, sumner2008} at degree 8 for \underline{s}tochastic \underline{tangle} (`stangle') invariants with mixed weight, since
it turns out that 
\[
g^{(8)}_{(51^3),(2^4),(2^4)} =1 \quad (\equiv g^{(8)}_{(2^4),(51^3),(2^4)}\equiv g^{(8)}_{(2^4),(2^4),(51^3)})\,.
\]
Thus there are three mixed weight stangle candidates, which would differ 
in the information they reveal about each leg of their ancestral star tree.

\end{appendix}


%
%
\vfill
\small
\end{document}